\documentclass[prl,aps,epsf,twocolumn,showpacs]{revtex4-1}
\usepackage{graphicx}
\usepackage{epsfig}
\usepackage{latexsym}
\usepackage{amsmath}
\usepackage{color}
\usepackage{array}

\newcommand{\e}{\varepsilon}
\renewcommand{\>}{\rangle}
\renewcommand{\(}{\left(}
\renewcommand{\)}{\right)}

\begin{document}

\title{Multichannel Generalization of Kitaev's Majorana End States and a Practical Route to Realize Them in Thin Films}

\author{Andrew C. Potter and Patrick A. Lee}
\affiliation{ Department of Physics, Massachusetts Institute of
Technology, Cambridge, Massachusetts 02139}

\date{\today}
\begin{abstract}
The ends of one-dimensional p+ip superconductors have long been predicted to possess localized Majorana fermion modes\cite{Kitaev}. We show that Majorana end states are robust beyond the strict 1D single-channel limit so long as the sample width is not much larger than the superconducting coherence length, and exist when an odd number of transverse quantization channels are occupied. Consequently the system undergoes a sequence of topological phase transitions driven by changing the chemical potential.  These observations make it feasible to implement quasi-1D p+ip superconductors in metallic thin-film microstructures, which offer 3-4 orders of magnitude larger energy scales than semiconductor-based schemes.  Some promising candidate materials are described.
\end{abstract}
\maketitle

Since the pioneering work by Kitaev\cite{Kitaev}, there has been growing interest in the condensed matter community to find a way to realize Majorana fermions in the laboratory.  Often described as ``half a fermion", Majorana fermions are expected to exhibit non-Abelian braiding statistics \cite{Read/Green,Ivanov} and have been proposed as a basis for topological quantum computers which would be protected from decoherence\cite{Kitaev,Nayak}.  While numerous schemes for realizing Majorana fermions have been proposed \cite{Moore/Read,Fu/Kane,Fujimoto,Sau10,Alicea10,PALee,Sau,Lutchyn,Oreg,Beenakker,Alicea}, up to now, all face significant technical challenges to implement.  The early proposal to use the $\nu = 5/2$ fractional quantum Hall state \cite{Moore/Read} has not yet received definite confirmation despite significant experimental efforts.  The proposal by Fu and Kane \cite{Fu/Kane} to use proximity induced superconductivity in the surface state of topological insulators opened up a promising new avenue but awaits further development in topological insulator materials.  

Recently, a number of proposals based on more traditional materials have appeared.  The basic idea is to use the Rashba-type spin-orbit coupling in combination with conventional $s$-wave superconductivity to produce an effective $p_x \pm ip_y$ 2D superconductor.  Magnetization is then introduced to remove one of the two components, leaving an effective $p_x + ip_y$ superconductor \cite{Fujimoto,Sau10,Alicea10,PALee} which is known to carry a single Majorana fermion in its vortex cores\cite{Read/Green,Ivanov}.  The Rashba coupling may be realized in semiconducting quantum wells \cite{Sau10,Alicea10}, non-centrosymmetric bulk superconductors, or on the surfaces and interfaces of strong spin-orbit coupled conventional superconductors\cite{Fujimoto,PALee}.  In one scheme a single component $p_x + ip_y$ superconductor would be realized by using the spin splitting of a ferromagnetic metal thin film\cite{PALee}.  Unfortunately, this requires control of the film thickness and interfaces to atomic precision in order to achieve an odd number of modes.  A more sophisticated scheme uses the fact that out-of-plane magnetization opens up a gap $\Delta_m$ at the Dirac crossing of a Rashba band.  As long as $\Delta_m$ is greater than the superconducting gap $\Delta$, an effective $p_x + ip_y$ superconductor results\cite{Fujimoto,Sau10,Alicea10}.  However, the electron density in such structures is extremely low\cite{Alicea10}, and exquisite purity will be needed to avoid carrier localization.

In addition to the above-mentioned challenges facing the construction of an effective $p_x + ip_y$ superconductor, realizing Majorana fermions as bound states in vortex cores
presents the additional difficulty of controllably creating and manipulating multiple superconducting vortices.  This has led to renewed interest in the original Kitaev idea\cite{Lutchyn,Oreg,Beenakker}, where Majorana particles are shown to be localized at the ends of a one-dimensional $p_x+ip_y$ superconducting wire.  In an interesting recent paper, this setup has been examined in great detail, including explicit plans to manipulate, braid, and measure these Majorana end states\cite{Alicea}.  

All previous discussions of Majorana end states, with one exception \cite{Comment_Beenakker}, assume that an effective one-dimensional band can be created by occupying only the lowest transverse sub-band of a quantum wire.  This again presents formidable technical challenges.  In this paper we move beyond the strict one-dimensional limit, and explore further the notion of Majorana end states in \emph{quasi}-one dimensional $p_x+ip_y$ superconductors.  We consider samples of finite width $W$, and allow the occupation of multiple transverse sub-bands.  Our key finding is that the Majorana end states are surprisingly robust.  They exist when an odd number of transverse channels are occupied, and are protected from low lying excitations as long as the sample width $W$ does not substantially exceed the superconducting coherence length $\xi_0 = \pi v_F/\Delta$.  Thus a series of quantum phase transitions between topological and conventional states occurs as a function of chemical potential.  Due to the condition $W < \xi_0$, the Majorana end states are stable over a range in chemical potential which exceeds $\Delta$.  We also find that these zero modes survive disorder, as long as the mean free path $\ell$ exceeds $W$.  

These results suggest that it is feasible to create \newline Majorana states in thin films.  To this end it is necessary to create a 2D $p_x + ip_y$ superconductor.  In the second half of this paper we propose a new scheme which combines elements of earlier proposals \cite{Fujimoto,Sau10,PALee}  and uses the surface states of a strong spin-orbit coupled metal with superconductivity and magnetization induced by proximity effect.  The advantage over semiconductor heterostructure schemes is that the energy scale of the Rashba splitting and the electron density are much higher.  Some promising candidate materials are described.

\textit{Majorana end states in quasi-one-dimensional wires - } We consider a rectangular $p_x+ip_y$ superconductor of length $L\gg \xi_0$, and width $W$.  For sufficiently large $W$, the system is fully two-dimensional and the low lying ($\e<\Delta$) states are chiral edge modes with a gap set by the finite circumference of the sample.  This gap is required by the fact that a single zero-energy Majorana state cannot exist in isolation, but can only occur in pairs.  The question is: how are these chiral edge states connected with the zero energy Majorana end states as the width decreases?  

To probe this crossover from 2D to 1D, we consider a square-lattice tight-binding model, $H = H_{t} + H_{\text{p-BCS}}$, of a single species of electrons with $p_x+ip_y$ BCS pairing:
\begin{eqnarray} H_{t} &=& \sum_{\<ij\>} -t\(c^\dagger_ic_j + \text{h.c.}\)-\sum_j \mu c^\dagger_jc_j \nonumber\\
H_{\text{p-BCS}} &=& \sum_j \Delta\(-ic^\dagger_{j+\hat{x}}c^\dagger_j+c^\dagger_{j+\hat{y}}c^\dagger_j\)+\text{h.c} \label{eq:TBModel}\end{eqnarray}
where $c^\dagger_j$ creates an electron on site $j$,  $t$ is the hopping amplitude, $\mu$ is the chemical potential, $\Delta$ is the p-wave pairing amplitude, and we work in units where the lattice spacing is unity. 

\begin{figure}[tttt]
\begin{center}
\hspace{-.2in}
\includegraphics[width=3in,height=1.5in]{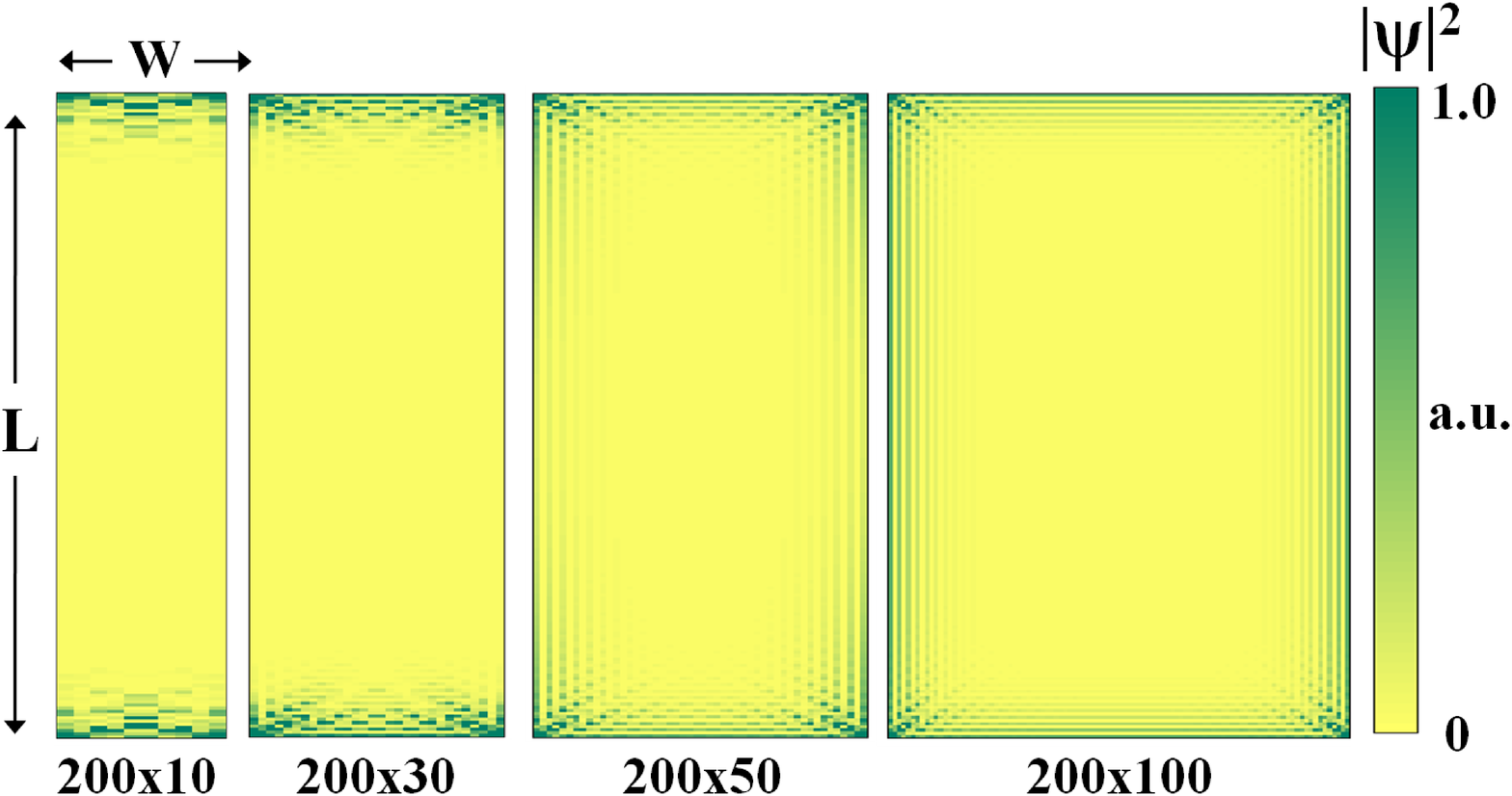}
\end{center}
\vspace{-.2in}
\caption{(Color online) $|\Psi|^2$ for zero-mode wave-functions at fixed coherence length: $\xi_0 \sim t/\Delta = 10$, chemical potential $\mu = -2t$, fixed sample length: $L=200$, and various sample widths:  $W =10$, $30$, $50$ and $100$ (from left to right).  These widths span the range from $W<\xi_0$, $W\sim \xi_0$, to $W\gg \xi_0$, and depict the crossover from the 1D regime with Majorana zero-modes localized at each end, to the 2D limit with chiral edge mode.}
\label{fig:WaveFns}
\end{figure}

Fig. \ref{fig:WaveFns} shows the lowest energy state wave-function, obtained by numerically diagonalizing Eq.(\ref{eq:TBModel}), for a sequence of samples with increasing widths.  For $W<\xi_0$, there are two well isolated Majorana end states localized at opposite ends. As $W$ is increased to $\gtrsim \xi_0$, these Majorana end states begin to spread along the edges of the sample with extent $\sim e^{W/\xi_0}$. For fixed $L$, these wave-functions eventually circumnavigate the sample to match up with the 2D edge mode picture for $W\gg \xi_0$.  The results of these simulations can be understood as follows: in the 2D limit, the edge modes are localized near the sample boundary and decay into the bulk with characteristic length $\xi_0$. As $W$ is decreased and approaches $\xi_0$, the tails of the edge mode wave-functions on opposite edges overlap and mix.  The mixing of these two counter-propagating edge modes forms a gap of order $e^{-W/\xi_0}$ along the mid-section of the sample, and squeezes the lowest energy state towards opposite ends of the sample.  
For sufficiently long samples ($L\gg e^{W/\xi_0}$) one finds spatially isolated zero-energy Majorana end states residing in opposite ends. 

\begin{figure}[tbbb]
\begin{center}
\hspace{-.2in}
\includegraphics[width=3in]{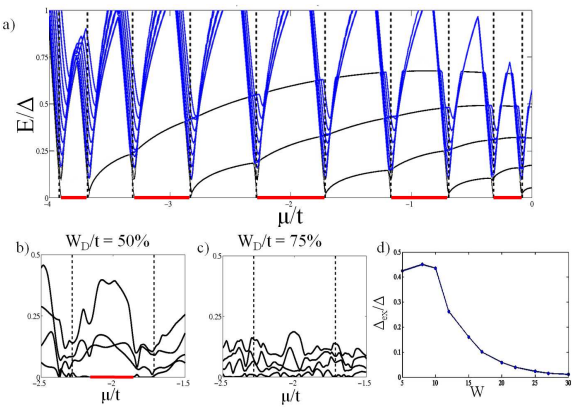}
\end{center}
\vspace{-.2in}
\caption{(Color online) a) Excitation spectrum as a function of chemical potential, $\mu$, for $100\times 10$ site lattice strip of $p_x+ip_y$ superconductor with $t/\Delta = 10 \sim \xi_0$. Blue lines are bulk excitations. Dashed lines denote the transverse confinement band-bottoms, which coincide with topological phase transitions between phases with Majorana zero-modes (shown in red), to topologically trivial gapped phases. b) and c) the effect of random on-site bulk disorder, normally distributed with width $W_D/t = 50\%$, and $75\%$ (left to right respectively). The Majorana zero-mode survives moderate disorder, but is destroyed for sufficiently strong disorder. Insert d) shows the exponential sensitivity of the excitation gap $\Delta_{\text{ex}}$ to $W$ for clean samples with $L=1000$, $t/\Delta = 5 \sim \xi_0$ and $\mu$ near $-2t$ (fine tuned to support a zero-mode).}
\label{fig:MuSweepAndDisorder}
\end{figure}

As in shown Fig. \ref{fig:MuSweepAndDisorder}d, the Majorana end modes are protected from low-energy excitations by a spectral gap that scales as $\sim \Delta e^{-W/\xi_0}$. Therefore, while a sample of any $W$ can in principle support Majorana end-states for sufficiently large $L\gg e^{W/\xi_0}$, in practice we have the restriction $W\lesssim \xi_0$ to avoid an exponentially small excitation gap. For $W < \xi_0$ and $L\gg \xi_0$, this gap is a sizeable fraction of the bulk superconducting gap, $\Delta$, and is largely insensitive to both $L$ and $W$. 


\textit{Topological phase transitions - }  In the quasi-1D regime, where $W \lesssim \xi_0$, the Majorana end states discussed above occur only when an odd number of transverse sub-bands is occupied.  We next examine the effects of changing chemical potential $\mu$.  The top panel of Fig.\ref{fig:MuSweepAndDisorder} shows the low lying excitation energies as a function of $\mu$ for an $L=100$, $W=10$ lattice with $t/\Delta = 10 \sim \xi_0$.  
The higher energy states shown in blue reside in the bulk and form two branches: excitations across the pairing gap and excitations to the next higher sub-band which respectively increase/decrease in energy with $\mu$. Due to the $p$-wave pairing symmetry, the pairing gap vanishes when $\mu$ coincides with a sub-band bottom (dashed lines).  

We now focus on low energy states lying within the bulk gap.  Starting from the lowest transverse sub-band and increasing $\mu$, one finds an alternating sequence of transitions between states with a non-degenerate zero-mode (shown in red) and doubly degenerate gapped states as described above.  The transition points coincide with $\mu$ passing through the bottom of a transverse sub-band, and are accompanied by a closing of the bulk gap (as is required for any topological phase transition).  

Intuitively one can imagine that each sub-band contributes a Majorana mode at each end of the sample.  An even number of Majorana fermions localized at a given end will mix and pair into full electron states at non-zero energy.  For an odd number, however, there is always one remaining unpaired Majorana mode per end. Thus, as a function of chemical potential $\mu$ the system undergoes a series of topological phase transitions between topological states with spatially isolated Majorana zero-modes to topologically trivial gapped states with no Majorana modes. We have verified this picture by checking that the non-degenerate (red) states have energy that is exponentially small in $L$, and that their wavefunctions can be written as $\psi=a+ib$, where $a$ and $b$ are real fermions localized in opposite ends.  On the other hand, the gapped states are doubly degenerate due to spatial parity symmetry $(x,y)\rightarrow -(x,y)$ and are complex fermions localized in the sample ends.


\textit{Robustness against disorder - }  In realistic experimental devices, some degree of random disorder is unavoidable.  To explore the effect of bulk disorder, we add a random onsite potential $H_{\text{Disorder}} = \sum_j V_j c^\dagger_j c_j$ to the tight-binding model (\ref{eq:TBModel}), where $\{V_j\}$ are normally distributed with variance $W_D$.  Fig. \ref{fig:MuSweepAndDisorder}b,c show a segment of the excitation spectrum as a function of $\mu$ for typical disorder realizations at two disorder strengths.  For weak to moderate disorder strength, the Majorana zero-mode survives, and the main effects of disorder are to narrow the spectral gap to the lowest excited state and to break the parity degeneracy of the excited states.  However, for sufficiently strong disorder, the excitation gap closes and the Majorana modes no longer exist for any $\mu$.  We have checked that the critical disorder strength depends on the relative size of the sample width and the mean-free-path $\ell \equiv  v_F\(\pi W_D^2 N(\mu)\)^{-1}$ (where $N(\e)$ is  the density of states, and $v_F$ is the Fermi velocity) with the Majorana mode surviving for $\ell \gtrsim W$.  It is important to note that, for moderate disorder strength, the existence of Majorana-modes and location of topological phase transitions are no longer precisely tied to the transverse quantization band-bottoms.  This suggests that the existence of Majorana zero-modes and alternating even-odd phase transition structure do not rely crucially on sharp transverse quantization.  

Another potentially important source of disorder is spatially varying edge orientation.  To assess its impact, we have also performed simulations for strips with meandering piece-wise-linear edges. We find similar results to those of bulk-disorder, with Majorana end-states surviving moderate spatial variations in $W$\cite{EPAPS}.

\begin{figure}[ttt]
\begin{center}
\includegraphics[width=3in]{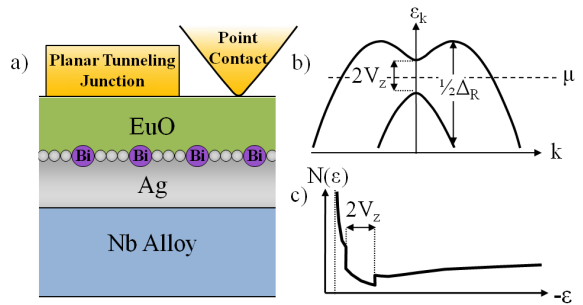}
\end{center}
\vspace{-.2in}
\caption{(Color online)Proposed metallic thin film based structure for creating a $p_x+ip_y$ superconductor (a).  A surface state with large Rashba splitting has been observed in the Ag(111) surface (gray circles) covered by 1/3 monolayer of Bi (purple circles)\cite{Dil}.  Adjacent layers of superconducting Nb alloy and ferromagnetic insulating EuO induce single-species $p_x+ip_y$ pairing in the surface state. Rashba and ferromagnetic split band structure (b) and corresponding  density of states, $N(\e)$ (c).  $N(\e)$ has a Van-Hove singularity at the Rashba band-bottom, and exhibits a dip due to the ferromagnetic magnetization gap $\Delta_m = 2V_z$. }
\label{fig:MaterialsStack}
\end{figure}

\textit{Thin-film realization - }    In order to realize the Majorana end-states discussed above, one must first construct a $p_x+ip_y$ superconductor.  As proposed in \cite{Fujimoto,Sau10,Alicea10}, an effective $p_x+ip_y$ superconductor can be engineered by combining three ingredients: 1) a 2D surface band with strong Rashba spin-orbit coupling term $H_{\text{R}} = \alpha_R {\bf\hat{z}} \cdot\({\bf\sigma}\times {\bf k}\)$,  2) s-wave spin-singlet pairing, and 3) a ferromagnetic splitting term, $H_{\text{FM}} = V_z\sigma_z$, to effectively remove one `spin-species'.  We have shown above that Majorana end states in quasi-1D $p_x+ip_y$ superconductors survive beyond the strict 1D limit and are robust against disorder.  These features make it feasible to realize strong Rashba coupling in the above recipe using metallic thin films rather than semiconductor heterostructures.

A particularly promising example is the surface state which forms when 1/3 monolayer of high Z elements such as Bi is grown on Ag(111) surface. This surface state is well isolated from bulk states and exhibits large Rashba coupling on the order of $\Delta_R \equiv 2\alpha_Rk_F = 4m^*\alpha_R^2 = 1.85$eV\cite{Dil}.  This value exceeds typical numbers achievable in semiconductor structures by \emph{three to four orders of magnitude}\cite{Alicea10}.  As discussed in \cite{Alicea10}, large $\Delta_R$ not only ensures that the induced s-wave pairing is effectively converted into $p_x+ip_y$ pairing, but also determines the maximum achievable carrier density in the topological superconducting phase.  These considerations lead us to propose the structure shown in Fig. \ref{fig:MaterialsStack}.

In this scheme, electrons confined to the Bi/Ag surface form a purely 2D Rashba split surface band.  Superconductivity can be induced by proximity to a conventional superconductor such as Nb.  Ferromagnetic splitting can be achieved by depositing a layer of ferromagnetic insulator, such as EuO, on top of the Bi/Ag surface layer.  Experiments on ultra-thin films of EuO have demonstrated sizeable ferromagnetic exchange energies equivalent to $V_z \sim $ $300$meV\cite{Moodera}.  While the actual induced Zeeman splitting will depend sensitively on the transparency of the EuO/Bi-Ag interface, even a small fraction of this previously demonstrated $V_z$ would be sufficient for our purposes. Planar tunneling could be used to characterize the surface Rashba and induced Zeeman energies.  Fig. \ref{fig:MaterialsStack}c shows the density of states $N(\e)$ for a surface band with Rashba and Zeeman terms.  $N(\e)$ has a square-root van Hove singularity at the Rashba band-bottom which has been observed directly in tunneling data\cite{Ast}.  $N(\e)$ also exhibits a dip due to the magnetization gap $\Delta_m = 2V_z$.  This dip is present only when the induced magnetization lies perpendicular to the surface, and could potentially be resolved by comparing tunneling data for in-plane and perpendicular magnetization. Finally, in order to induce single-species $p_x+ip_y$ pairing, $\mu$ must be tuned to lie within the Zeeman gap (see Fig. \ref{fig:MaterialsStack}b).  This tuning can be achieved, for example, by electrostatic gating.  Local tunneling experiments can be used to unambiguously verify the presence of Majorana modes. These would show up as zero-energy mid-gap states, and at sufficiently low temperature would exhibit quantized conductance of $G = 2e^2/h$ for resonant Andreev reflection\cite{MIRAR}.

We have also carried out simulations of finite strips of spin-full fermions in the presence of Rashba, Zeeman and s-wave pairing terms.  The results agree with those shown above, and display the alternating appearance and disappearance of Majorana fermions with $\mu$, as long as $\mu$ lies within the Zeeman gap.  

\textit{Conclusions and Discussion - } We have shown that robust Majorana end states in 1D $p_x+ip_y$ superconductors survive beyond the strict 1D limit, so long as the sample width $W$ is not significantly larger than the superconducting coherence length $\xi_0$, opening the door for realizing Majorana fermions in metallic thin film microstructures with large Rashba spin-orbit coupling.  Such materials offer orders of magnitude larger energy scales and carrier concentrations compared to analogous semiconductor-based proposals.  The Majorana states can be controlled either by fine-tuning $\mu$ with electrostatic gates, or by controlling domain walls in the Ferromagnetic insulator.  For example, an effective $p_x+ip_y$ region can be created at a domain wall separating two regions of opposite in-plane magnetization. Our results raise the exciting possibility that Majorana fermions could be realistically implemented in microstructures defined by conventional lithography techniques, and could be efficiently manipulated and braided simply by electrostatic gating or other means. 
 
\textit{Acknowledgements - } We thank Jason Alicea for many helpful discussions and thank Hugo Dil, Roland Kawakami and Jagadeesh Moodera for discussions on the materials aspect.  This work was supported by DOE DE--FG02--03ER46076 (PAL) and NSF IGERT DGE-0801525 (ACP).

\newpage
{\bf Appendix A: Simulations with a Meandering Edge: }
In addition to random bulk on-site disorder, one may also worry about the effects of spatially varying non-parallel edges.  This type of edge disorder mixes transverse sub-bands by edge-scattering, and if sufficiently strong may close the bulk gap and destroy the Majorana end-states.  However, we find that the Majorana end-states are robust against moderate spatial variations in edge width and orientation.

To demonstrate the robustness of the multi-channel scheme to moderate amounts of edge-variance, we have conducted simulations of $p+ip$ strips with meandering edge profiles.  Random meandering edges were created by introducing a strong electrostatic confinement potential $V_{\text{conf}}(x,y)$, which is zero inside the strip and transitions to $V_{\text{conf}}(x,y)=4t$ outside the strip.  The top and bottom edges were independently formed by dividing the length of the strip into uniformly random intervals of average length $L_{\text{meander}}$.  The position of the top and bottom edges were chosen randomly at the endpoint of each of the aforementioned intervals with average $\pm W/2$ and Gaussian spread $\sigma_W$ .  In between these end-points, the top and bottom edge profiles were linearly interpolated (in discrete steps).  Finally, in order separate the effects of non-parallel edges from edge jagged-ness, $V_{\text{conf}}(y)$ was smoothed over the length scale of one lattice spacing by applying a Gaussian filter along y.

\begin{figure}[ttt]
\begin{center}
\includegraphics[width=3.5in]{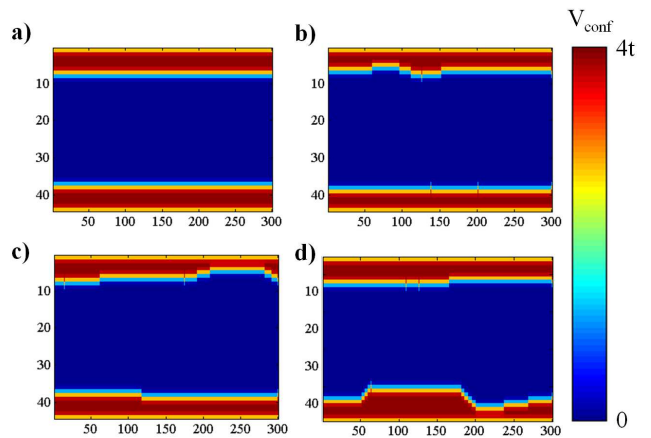}
\end{center}
\vspace{-.2in}
\caption{(Color online) Meandering edge confinement potential, $V_{\text{conf}}$, spatial profiles for 300$\times$30 strip with $t=30$, $\Delta=1$.  (a)--(d) show random samples with fixed meandering correlation length: $L_{\text{meander}} = 60 = 3W$, and  increasing width variance: $\sigma_{W}$ = 0, 1, 1.5, and 2 respectively. Corresponding excitation spectra shown in Fig. \ref{fig:MeanderingEdgeSpectra} below. }
\label{fig:MeanderingEdgeProfiles}
\end{figure}

\begin{figure}[ttt]
\begin{center}
\includegraphics[width=3.5in]{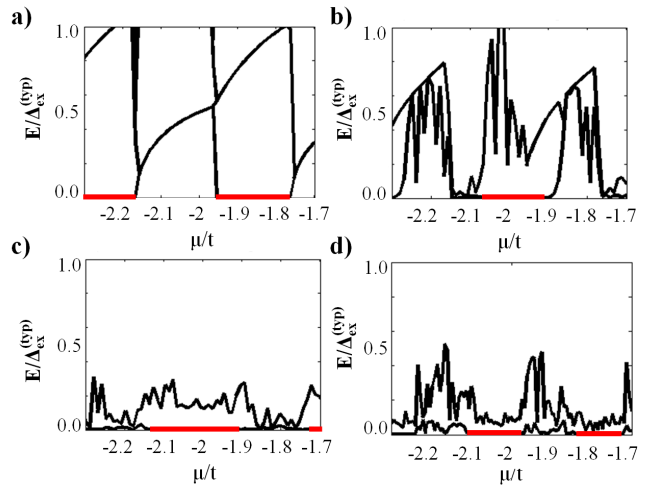}
\end{center}
\vspace{-.2in}
\caption{(Color online) Corresponding excitation spectrum as a function of $\mu$, for meandering edge samples shown in Fig. \ref{fig:MeanderingEdgeProfiles}(a)--(d) above.  Excitation energies are measured with respect to the typical excitation energy $\Delta_{\text{ex}}^{(\text{typ})}=0.1\Delta$ for the straight-edge case (a).  The effects of the smooth edge disordering are similar to those of bulk onsite disorder:  increasing $\sigma_W$ decreases the bulk excitation gap by mixing transverse sub-channels.  However, for moderate $\sigma_W$ the Majorana end-states survive and continute to appear and dissapear in an alternating fashion as $\mu$ is swept.}
\label{fig:MeanderingEdgeSpectra}
\end{figure}

Representative simulations of samples with meandering edges are shown in Figures \ref{fig:MeanderingEdgeProfiles} and \ref{fig:MeanderingEdgeSpectra}.  Fig. \ref{fig:MeanderingEdgeProfiles} a.--d. show colormaps of $V_{\text{conf}}$ for increasing edge variance: $\sigma_{W}$ = 0, 1, 1.5, and 2 lattice spacings respectively.  These variances represent fractional variances in the sample width of 0\%, 6.7\%, 10\%, and 13.3\% respectively. 

Fig. \ref{fig:MeanderingEdgeSpectra} shows the corresponding excitation spectra, as a function of chemical potential $\mu$.  In these figures, energies are measured with respect to typical straight-edge excitation gap scale of $\Delta_{\text{ex}}^{(\text{typ})}=0.1\Delta$ shown in (a).  Like bulk-disorder, the effect of edge disorder is to reduce the bulk excitation gap, with larger variance in the edge geometry (either through increasing $\sigma_W$ or decreasing $L_{\text{meander}}$) leading to stronger reduction.  

Crucially, the Majorana zero-modes (highlighted in red) survive moderate amounts of edge-variance.  Furthermore, the alternating sequence of topological phase transitions as a function of $\mu$ survives with meandering edges.  However, as with bulk disorder, the locations of the phase transitions are no longer tied to the clean transverse sub-band bottoms.  The amount of edge-variance that can be tolerated depends on the system geometry ($W$, $L$, and  $L_{\text{meander}}$) and coherence length $\xi_0$.  From simulations of samples with various widths (but with fixed $\xi_0/W \sim 1$), we find that typically, the Majorana end-states can survive up to $\sim 15\%$ variation in the edge width. 
\newpage 


\end{document}